\documentclass[epsfig,floats,prl,twocolumn,preprintnumbers,floatfix,superscriptaddress]{revtex4-1}
\usepackage{graphicx}
\usepackage[latin2]{inputenc}
\usepackage{amsmath}
\usepackage{amssymb}
\usepackage{bm}
\usepackage{color}

\begin{document}
\title{Gravity Governs Shear Localization in Confined Dense Granular Flows}
\author{M.\ Reza Shaebani}
\email{shaebani@lusi.uni-sb.de}
\affiliation{Department of Theoretical Physics and Center for Biophysics, 
Saarland University, 66123 Saarbr\"ucken, Germany}
\author{J\'anos T\"or\"ok}
\affiliation{MTA-BME Morphodynamics Research Group, Department of Theoretical 
Physics, Budapest University of Technology, and Economics, Budapest H-1111, 
Hungary}
\author{Maniya Maleki}
\affiliation{Department of Physics $\&$ Optics Research Center, Institute for 
Advanced Studies in Basic Sciences, Zanjan 45137-66731, Iran}
\author{Mahnoush Madani}
\affiliation{Department of Physics $\&$ Optics Research Center, Institute for 
Advanced Studies in Basic Sciences, Zanjan 45137-66731, Iran}
\author{Matt Harrington}
\affiliation{Department of Physics and Astronomy, University of Pennsylvania, 
Philadelphia, Pennsylvania 19104, USA}
\author{Allyson Rice}
\affiliation{Department of Biophysics Institute, UT Southwestern Medical 
Center, Dallas, TX 75390, USA}
\author{Wolfgang Losert}
\affiliation{Department of Physics, University of Maryland, College Park, 
Maryland 20742, USA}

\begin{abstract}
Prediction of flow profiles of slowly sheared granular materials is a 
major geophysical and industrial challenge. Understanding the role of 
gravity is particularly important for future planetary exploration in 
varying gravitational environments. Using the principle of minimization 
of energy dissipation, and combining experiments and variational analysis, 
we disentangle the contributions of the gravitational acceleration and 
confining pressure on shear strain localization induced by moving fault 
boundaries at the bottom of a granular layer. The flow profile is 
independent of the gravity for geometries with a free top surface. 
However, under a confining pressure or if the sheared layer withstands 
the weight of the upper layers, increasing gravity promotes the transition 
from closed shear zones buried in the bulk to open ones that intersect the 
top surface. We show that the center position and width of the shear zone 
and the axial angular velocity at the top surface follow universal scaling 
laws when properly scaled by the gravity, applied pressure, and layer 
thickness. Our finding that the flow profiles lie on a universal master 
curve opens the possibility to predict the quasistatic shear flow of 
granular materials in extraterrestrial environments.
\end{abstract}

\maketitle

Superficial layers of many planets and asteroids are covered by granular 
materials that range in size from dust and sand to gravel and boulder \cite{Hestroffer19}. 
These layers result from several processes, such as volcanic activity, fragmentation, 
and erosion, and extend to variable depths. Although mechanical behavior of granular 
matter in low-gravity conditions has recently attracted attention \cite{Harth18,
Murdoch13,Sack15,Katsuragi18,Noirhomme18,Murdoch13b}, it is generally unclear how 
various geological processes that occur in near-surface granular layers of celestial 
bodies depend on the gravitational acceleration $g$. Particularly, the localization 
of shear in a slow granular flow can initiate more catastrophic phenomena such as 
avalanches, earthquakes and faulting \cite{Scott96,Daerr99,Oda98}. Despite intensive 
studies on how and where strain is localized in shear zones \cite{Hartley03,Losert00,
Mueth00,Kollmer20,Schall10,Moosavi13,Knudsen09,Unger04,Fenistein04,Fenistein03,Cheng06,
Fenistein06}, the mechanism of shear band formation in slow granular flows (thus 
the role of $g$) cannot be described by a strain-stress constitutive law, as the 
flow profile and stress are independent of the shear rate. However, it is crucial 
to predict the shear flow of granular matter in varying gravitational environments 
to understand the evolution of planetary and asteroid surfaces and for a successful 
design of planetary and asteroid exploration programs and operation of landers, 
rovers, and sub-surface sampling devices.

\begin{figure}[t]
\centerline{\includegraphics[width=0.46\textwidth]{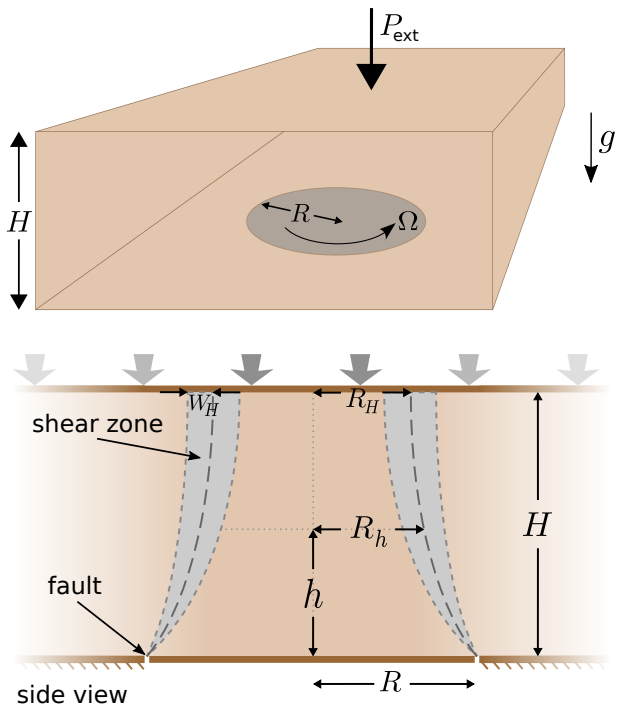}}
\caption{Sketch of the shear localization zone created by the rotation of a 
circular region at the bottom of a granular pile. The shear zone reaches the top 
surface of relatively shallow layers if the external pressure $P\!\!_\text{ext}$ 
is much weaker than the mean hydrostatic pressure $\frac12 \rho g H$ induced by 
gravity; otherwise the zone is buried in the bulk.}
\label{Fig1}
\end{figure}

In granular flows created by slowly shearing the boundaries, the strain often 
localizes in narrow regions of the order of a few particle diameters near the 
moving boundary \cite{Mueth00,Losert00}. Nevertheless, wide shear zones can 
also be created when a shear fracture \cite{Riikila17} propagates in the bulk 
away from the boundaries, and the sides of the fault move past each other in 
deep layers--- the process that is of particular importance in geology; see 
Fig.\,1 for an example of a circular fault line in the bulk. In 
experiments, wide shear zones have been generated by pinning the shear band 
in the bottom of Couette geometries far from the cylinder wall \cite{Fenistein03,
Fenistein04,Cheng06,Fenistein06}. The position and horizontal extent of the 
resulting shear zone have shown power-law scalings with the height $H$ of the 
pile \cite{Fenistein03,Fenistein04}; the zone evolves towards the axis of 
cylinder and grows wider as it takes the advantage of the gravity to reach 
the top surface. However, a transition occurs in the flow structure from 
shear zones that intersect the top surface to closed ones buried in the bulk 
as $H$ is further increased \cite{Unger04,Cheng06,Fenistein06,Torok07}. 

\begin{figure*}[t]
\centering
\includegraphics[width=0.8\textwidth]{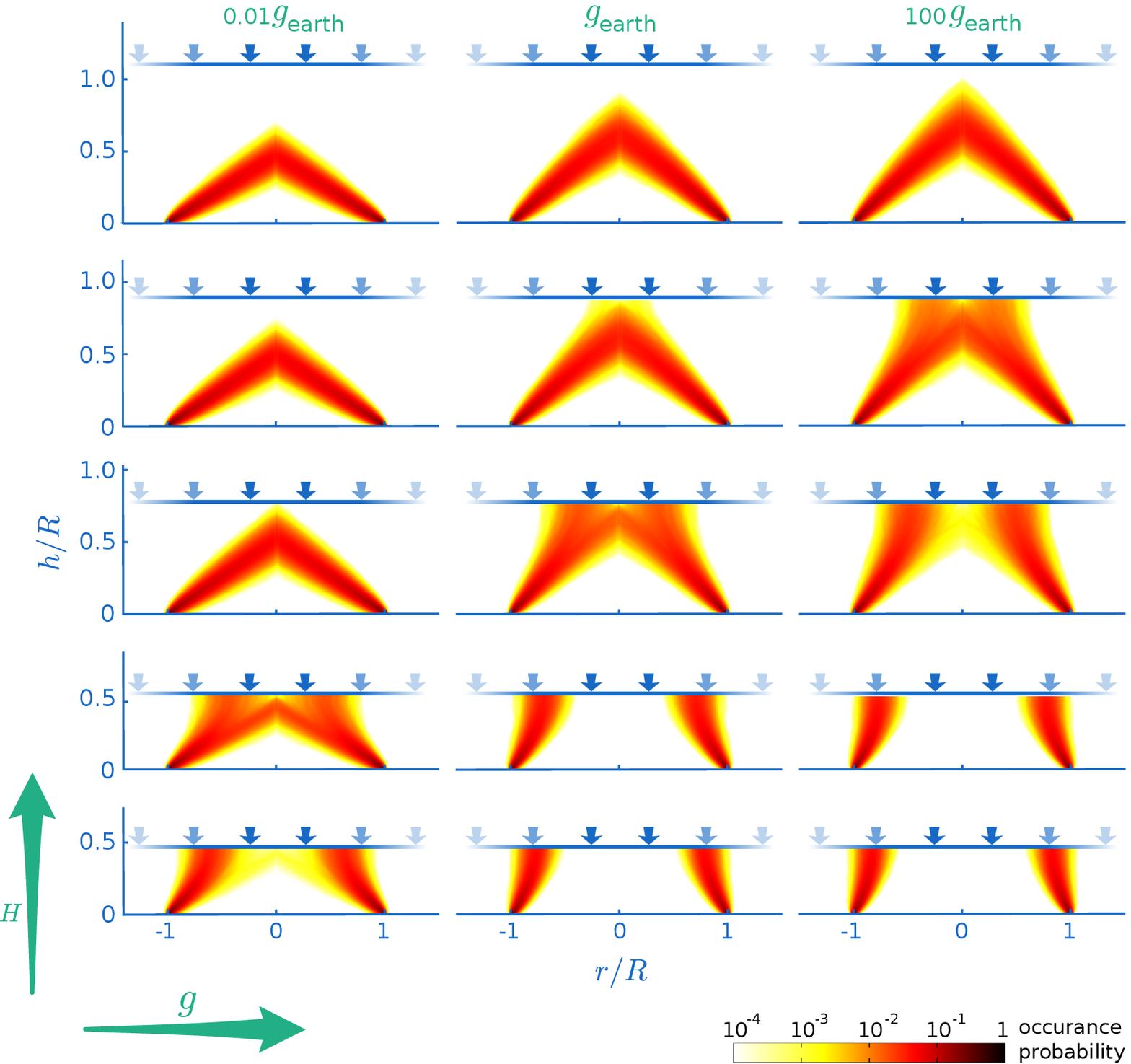}
\caption{Dependence of the shear-zone type on gravitational acceleration 
and filling height. The filling height is $H{/}R{=}0.48$, $0.55$, $0.77$, 
$0.88$ and $1.10$ from the bottom to top row, respectively. The gravitational 
acceleration in left, middle, and right panels is $\widetilde{g}{\equiv}g{/}
g_\text{earth}{=}0.01$, $1$ and $100$, with $g_\text{earth}$ being the surface 
gravity of the earth. In all panels, $P\!\!_\text{ext}$ equals the reference 
pressure $P\!\!_\circ$; see main text. The color identifies the occurrence 
probability at the given point. Open, closed, and combined shear zones can 
be distinguished.}
\label{Fig2}
\end{figure*}

While decreasing gravity insignificantly influences the flow profile near walls 
\cite{Murdoch13,Murdoch13b}, the profile away from the boundaries was shown 
to be sensitive to the choice of $g$ for soft particles \cite{Singh15}. Upon 
varying $g$ in shearing of rigid particles, one expects that the shear rheology 
of packings without a constraining top boundary would not be affected as $g$ 
only rescales the pressure gradient \cite{Unger04}; in contrast, under a 
confining pressure or if the sheared layer withstands the weight of upper 
layers, the relative pressure difference between successive horizontal layers 
of grains depends on $g$, suggesting that the gravity alters the shear flow 
profile in this case. 

We combine numerical analysis and experiments to explore the role of gravitational 
acceleration $g$ in slow shear flows of granular materials. We consider a rotating 
circular fault at the bottom of a horizontally infinite granular layer under 
confining pressure $P\!\!_\text{ext}$ (Fig.\,\ref{Fig1}) and derive the shape 
of the shear localization zone from the variational principle of least energy 
dissipation \cite{Unger04,Torok07}. Increasing gravity extends the open shear 
zone regime to thicker layers and pushes the zone away from the axis of rotation 
while slightly decreases its width. For closed shear zones, a stronger gravity 
pulls the shear zone towards the top surface and increases the extent of it. 
Although the flow profile is controlled by both $g$ and $P\!\!_\text{ext}$, we 
verify that they are not interchangeable and show that the position and width 
of the shear zone and the axial angular velocity at the top surface lie on 
universal master curves when properly scaled by $g$, $H$, and $P\!\!_\text{ext}$. 
We also carry out experiments in a Couette cell geometry to validate the numerical 
findings. Our results enable us to predict the shear flow profile in slow granular 
flows for a given set of filling height, confining pressure, and gravitational 
acceleration.  

\section{Results}
{\bf Shear flow profile from energy dissipation minimization.}
In order to clarify the role of gravity on the strain localization in sheared 
granular matter, we consider a shear band initiated by a rough disk with radius 
$R$ rotated at angular velocity $\Omega$ at the bottom of a horizontally infinite 
granular layer of height $H$ as shown in Fig.\,1 (becoming two infinite 
half-plates moving past each other in the limit $R{\rightarrow}\infty$). Here we 
assume the same effective friction coefficients in the bulk and near the boundaries, 
thus, the flow profile remains independent of the choice of $\mu$ (though, in 
general, friction plays a major role in stress transmission in granular materials 
\cite{Shaebani07,Goldenberg05,GDRMiDi04,Shaebani09}). To describe the flow profile, 
we apply the principle of least dissipation proposed by Onsager for irreversible 
time-independent phenomena \cite{Onsager31,Onsager31b,Baker78}. The validity of 
this variational approach has been previously confirmed by accurately predicting 
the strain localization path in various sheared granular systems \cite{Moosavi13,
Unger04,Torok07,Unger07,Knudsen09,Szabo14,Borzsonyi09}. 

We require a stationary flow that minimizes the rate of energy dissipation and 
matches the boundary constraints. Denoting the radial coordinate of the shear 
band at height $h$ with $R_h$ and taking the cylindrical symmetry of the geometry 
into account, the variational problem is traced back to finding an optimal $R_h$ 
function that satisfies $R_0{=}R$ while the other boundary at $h{=}H$ is free. 
In the narrow shear band approximation \cite{Unger04}, the dissipation rate 
is given by the shear stress $\sigma_\text{tn}\!=\!\mu\,\big(\rho\,g\,(H{-}h){+}
P\!\!_\text{ext}\big)$ times the sliding velocity between the two sides $R_h
\,\Omega$ integrated over the whole shear band. For the shear stress along the 
yielding surface we use Coulomb friction between two sliding bodies and assume 
that the Janssen effect plays no role due to the continuous agitations induced 
by collisions and slip events which causes a slight creep in the system. Up to 
a constant prefactor, the expression to be minimized can be formulated as 
\begin{equation}
\displaystyle\int_0^H \!\!\!\!\!\! R_h^2 \, \displaystyle\sqrt{1{+}
\Big(\frac{\text{d}R_h}{\text{d}h}\Big)^2} \,\, \Big(1{-}\frac{h}{H}{+}
\frac{P\!\!_\text{ext}}{\rho\,g\,H}\Big)\,\text{d}h=\text{min},
\label{Eq:VM1}
\end{equation}
thus, the optimal path is generally determined by $g$, $P\!\!_\text{ext}$, $H$, 
and $R$. Yet, the variational approach predicts that the path is independent of 
$g$ and $P\!\!_\text{ext}$ and solely determined by $H$ and $R$ in two extreme 
limits: For a very weak gravity or strong $P\!\!_\text{ext}$, $1{\ll}
\frac{P\!\!_\text{ext}}{\rho\,g\,H}$ and the fabric is isotropic under the 
homogeneous pressure \cite{Shaebani09b}. In this regime, equation\,(\ref{Eq:VM1}) 
reduces to $\int\!\!_{_{\!\!0}}^{^{\,H}} R_h^2 \, \sqrt{1{+}(\frac{\text{d}
R_h}{\text{d}h})^2}\,\text{d}h{=}\text{min}$, which leads to 
$h\,{=}\!\int_R^{R_h}\!\!\!\frac{\text{d}r}{\sqrt{r^4{/}A^2{-}1}}$ by solving 
the Euler-Lagrange equation ($A$ is a constant). The solution can be expressed 
in terms of the hypergeometric function ${}_{2}F_{1}$ as 
$h{=}r\sqrt{r^4{/}A^2{-}1}\,{}_{2}F_{1}(\frac34,1;\frac54;r^4{/}A^2)$ with 
the boundary condition $r(h{=}0){=}R$. On the other hand, in the extreme limit 
of strong gravity or if $P\!\!_\text{ext}{\rightarrow}0$ (as for a layer with 
a free surface), $\frac{P\!\!_\text{ext}}{\rho\,g\,H}{\ll}1$ and the strong 
pressure gradient leads to the variational problem $\int\!\!_{_{\!\!0}}^{^{\,H}} 
R_h^2\, \sqrt{1{+}(\frac{\text{d}R_h}{\text{d}h})^2}\,\,(1{-}\frac{h}{H})\,
\text{d}h{=}\text{min}$ \cite{Unger04}, which is again independent of $g$ 
and $P\!\!_\text{ext}$.

The above optimization procedure leads to an instantaneous narrow shear band. 
However, the material strength is practically affected in the vicinity of yield 
events \cite{Shaebani08} and the resulting random structural changes slightly 
vary the minimal path in the next instance. In a fluctuating-band version of 
the model \cite{Torok07}, we introduce local fluctuations of the path around 
the current shear band (see {\it Methods} section for details). The resulting 
shear zone gains a finite width in the course of such a self-organized process 
where the global optimum path itself modifies the medium in which the optimization 
is carried out. We obtain the shear profile by ensemble averaging over all 
instantaneous shear bands calculated via the optimization equation\,(\ref{Eq:VM1}) 
for a given geometry $H{/}R$ and a set of $g$ and $P\!\!_\text{ext}$ parameters. 

\noindent{\bf Gravity governs the strain localization profile.}
We perform extensive numerical simulations based on the optimization scheme to 
clarify the individual roles of $H$, $g$ and $P\!\!_\text{ext}$ on the shear 
flow profile as they affect the variational problem differently; $H$ influences 
both the integral limit and integrand in equation\,(\ref{Eq:VM1}). Also, $g$ 
and $P\!\!_\text{ext}$ impose a pressure gradient or an isotropic pressure, 
respectively, thus they are not interchangeable factors.

In case of a free top surface, i.e.\ $P\!\!_\text{ext}{=}0$, we checked that the 
flow profile is controlled by $H{/}R$ \cite{Fenistein03,Fenistein04,Unger04} 
and that gravity plays no role. However, the strength of the pressure gradient 
induced by $g$ is expected to influence the flow profile for $P\!\!_\text{ext},
{\neq}0$ since an isotropic confining pressure changes the micro-structure and 
dilation behavior of the material \cite{Murdoch13b,Singh15,Bandi19,Kobayakawa18,
Shaebani12,Das19}. To verify this expectation, we fix the applied pressure at a nonzero 
value and vary $g$ by several orders of magnitude at different values of $H$ 
(For ease of comparison, we introduce a reference pressure level $P\!\!_\circ$ 
that a pile with height $H{/}R{=}0.1$ consisting of glass beads of size $d$ 
applies on the bottom disk with $R{/}d{=}100$ in the gravitational acceleration 
of the earth $g_\text{earth}$). Interestingly, by increasing the gravity, the 
shear zone reaches the top surface for deeper layers as shown in Fig.\,2. 
It can be also seen that the gravity pushes the open shear zones away from the 
axis of rotation and slightly reduces the extent of the zone. This is in contrast 
to the closed zones, where the gravity even increases the extent of the strain 
localization zone while pulls it towards the top surface. 

\begin{figure}[t]
\centering
\includegraphics[width=0.47\textwidth]{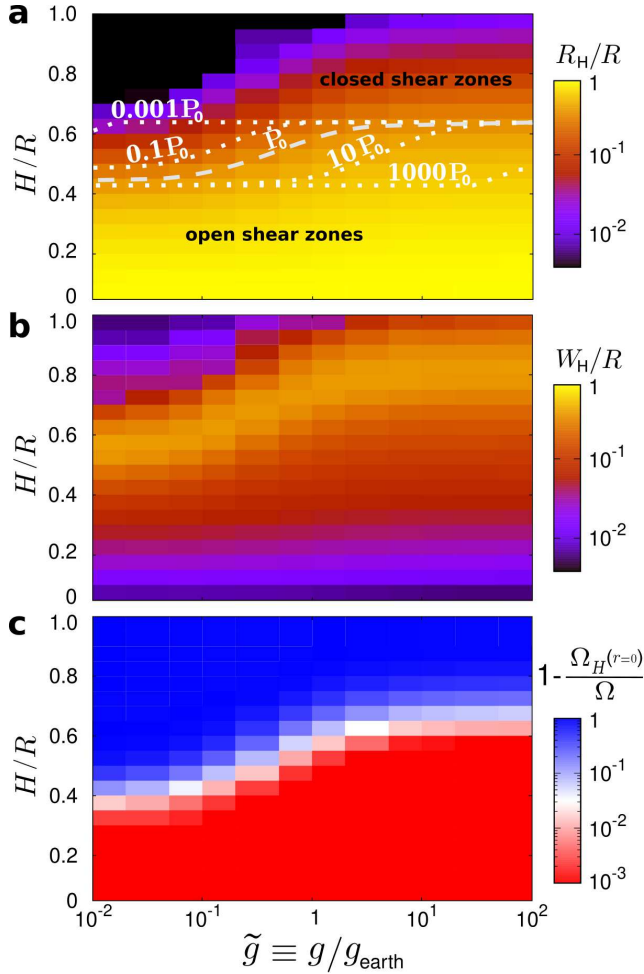}
\caption{Phase diagrams of the shear-zone position and width and the axial 
angular velocity at the top surface. {\bf a}, Radial coordinate $R_H$ of the center 
of shear zone (scaled by the radius $R$ of the bottom disk) in the ($g$, $H$) space, 
for $P\!\!_\text{ext}{=}P\!\!_\circ$. The dashed line denotes the threshold at which 
the open shear zone eventually reaches the rotation axis. The dotted lines represent 
the same threshold for other values of $P\!\!_\text{ext}$. {\bf b}, Scaled width of 
the shear zone at the top surface, $W_H{/}R$. {\bf c}, Deviation of the axial angular 
velocity $\Omega_H(r{=}0)$ at the top surface from the driving rate $\Omega$ of the 
bottom disk.} 
\label{Fig3}
\end{figure}

We quantify the center of the shear zone at a given height $h$ as the radial 
distance $R_h$ at which the rate of the strain reaches its maximum, i.e.\ 
$\text{d}\epsilon_{r\theta}{/}\text{d}r|_{_{r{=}R_h}}{=}0$. The width $W_h$ 
is taken to be the variance of the shear rate $\text{d}\Omega_h(r){/}\text{d}r$ 
around the peak at $r{=}R_h$. Figure\,3 summarizes the behavior of 
the center position $R_H$, width $W_H$, and axial angular velocity $\Omega_H(r{=}0)$ 
at the top surface in the ($g$, $H$) space. The continuous transition from open 
to closed shear zone along the $H$-axis \cite{Fenistein06,Cheng06} manifests 
itself e.g.\ in the abrupt change in the axial angular velocity from $\Omega$ 
to $0$ (within the numerical accuracy of the measurements); see Fig.\,3c. 
The gravity enhances the open shear zone regime to higher values of $H$. We 
can identify the transition from open to closed shear bands as, for example, 
when the shear localization region reaches the rotation axis. As shown in 
Fig.\,3a, the transition line in the $g$-$H$ phase diagram depends 
on the applied pressure $P\!\!_\text{ext}$ but approaches a saturation level 
towards each extreme limit of $g$. 

\noindent{\bf Universal characteristics of shear zones.} 
The radial coordinate of the shear band at the top surface, $R_H$, monotonically 
increases with $g$ for any choice of $H$ and $P\!\!_\text{ext}$; see Fig.\,4a. 
Denoting the scaled gravity and pressure by $\widetilde{g}{\equiv}g{/}
g_\text{earth}$ and $\widetilde{P}{\equiv}P\!\!_\text{ext}{/}P\!\!_\circ$, 
we introduce a dimensionless parameter $\lambda{=}\widetilde{g}{/}\widetilde{P}$ 
with which we can tune the relative strength of the gravity compared to the 
applied pressure. Interestingly, when plotting $R_H$ versus $\lambda$ in 
Fig.\,4b, we achieve a striking data collapse for each filling 
height. The universal master curve at a given $H$ suggests that to predict 
the flow profile as a function of gravity, one can instead use the inverse of the applied 
pressure as control parameter. To validate this hypothesis we carried out shear experiments 
in a modified Couette cell with a transparent plate on the top of the granular 
layer to apply pressure (see Fig.\,5a and {\it Methods} section for 
details). Figure~4b shows that for several different filling heights 
the experimental data (obtained by varying the mass of the top plate) follow 
the master curves satisfactorily. Following the power-law scaling that was 
previously reported in cells without confining pressure \cite{Fenistein03,Fenistein04}, we 
find that the entire data for all $g$, $P\!\!_\text{ext}$, $H$, and $R_H$ 
values lie on a universal curve
\begin{equation}
1-\displaystyle\frac{R_H}{R}=f(\lambda)\Big(\displaystyle\frac{H}{R}\Big)^{5{/}2},
\label{Eq:RHR}
\end{equation}
where $f(\lambda)$ is a logistic function of $\text{ln}\lambda$ (see inset of 
Fig.\,4c):
\begin{equation}
f(\lambda)=a+\displaystyle\frac{b}{1{+}\text{e}^{-(\ln\lambda{+}c)}}\,,
\label{Eq:logistic}
\end{equation}
with $a$, $b$, and $c$ being fit parameters.

\begin{figure*}
\centering
\includegraphics[width=0.99\textwidth]{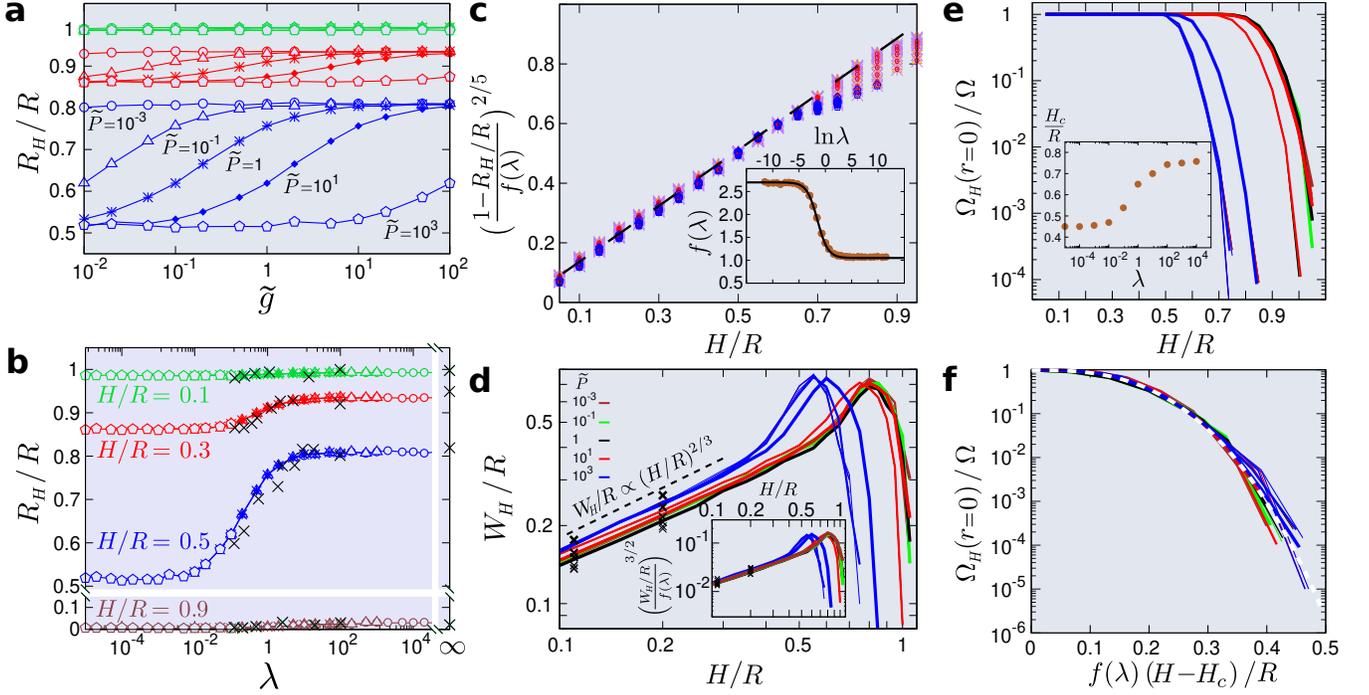}
\caption{Universal characteristics of shear zones. {\bf a}, Radial coordinate 
of the shear band at the top surface, $R_H$, in terms of the scaled gravitational 
acceleration $\widetilde{g}{\equiv}g{/}g_\text{earth}$. Different filling heights 
(applied pressures $\widetilde{P}{\equiv}P\!\!_\text{ext}{/}P\!\!_\circ$) are 
indicated with different colors (symbols). {\bf b}, $R_H$ vs $\lambda{=}\widetilde{g}
{/}\widetilde{P}$. The data shown in panel ({\bf a}) collapses on a universal master 
curve for each $H$. Crosses represent experimental data. {\bf c}, Scaled shear-zone 
position via equation~(\ref{Eq:RHR}) vs $H{/}R$. The dashed line indicates equality. 
Inset: $f(\lambda)$ vs $\ln\lambda$. The solid line is given by equation~(\ref{Eq:logistic}) 
with $a{\simeq}2.7$, $b{\simeq}-1.7$, $c{\simeq}1.5$. {\bf d}, Width of the shear 
zone in terms of $H$ for various $P\!\!_\text{ext}$ (different colors) and $g$ 
($\widetilde{g}$ grows from $0.01$ to $100$ with increasing line thickness). 
Crosses represent experimental data at $H{/}R{\simeq}0.11$, $0.20$ (increasing 
symbol thickness with $P\!\!_\text{ext}$). Inset: Collapse of the data onto a master 
curve using equation~(\ref{Eq:WHR}). {\bf e}, Axial angular velocity at the top 
surface (scaled by the driving rate $\Omega$ of the bottom disk) as a function 
of $H$. Same colors and line thicknesses as in panel ({\bf d}). The inset shows 
the variation of the cutoff height $H_c$ with $\lambda$. {\bf f}, Decay of 
$\Omega_H(r{=}0)$ with the scaled excess filling height $\widetilde{H}=f(\lambda)
\,(H{-}H_c){/}R$. The white dashed line is the scaling form given by 
equation~(\ref{Eq:OmegaH}).}
\label{Fig4}
\end{figure*}

The width $W_H$ of the shear zone at the top surface, shown in Fig.\,4d, 
follows a scaling law  
\begin{equation}
\displaystyle\frac{W_H}{R}\propto f(\lambda)\Big(\displaystyle\frac{H}{R}
\Big)^{\beta},
\label{Eq:WHR}
\end{equation}
with the exponent $\beta{=}2{/}3$, indicating that the extent of the shear 
zone grows faster than diffusively but slower than linearly with $H$. A 
similar growth exponent was observed in open surface shear experiments 
\cite{Fenistein03,Fenistein04}. There are, however, growing deviations 
from the scaling behavior as the filling height approaches the transition 
threshold to the closed shear zones. While the exponent $\beta$ remains 
unchanged, increasing $g$ (or decreasing $P\!\!_\text{ext}$) slightly 
decreases $W_H$. It was shown that gravity causes decreasing amplitude 
of grain displacement in signal propagation which slows down the 
dispersion of signal width \cite{Hong01}. The inset of Fig.\,\ref{Fig4}d 
shows that a data collapse is achieved when rescaling according to 
equation\,(\ref{Eq:WHR}) using a free fit to the logistic 
function\,(\ref{Eq:logistic}).

In shallow layers, the top surface directly above the circular fault region 
rotates with the driving rate $\Omega$ of the bottom disk. As $H$ increases, 
the angular velocity $\Omega_H(r)$ at the top surface gradually reduces 
during the crossover from open to closed shear zones. It can be seen from 
Fig.\,\ref{Fig4}e that the cutoff height $H_c$ (at which the axial angular 
velocity at the top surface $\Omega_H(r{=}0)$ starts decreasing) depends 
on $P\!\!_\text{ext}$ and $g$. By subtracting $H_c$ from the total height, 
we introduce a scaled excess filling height $\widetilde{H}{=}f(\lambda)
\displaystyle\frac{H{-}H_c}{R}$. Then, the angular velocity data remarkably 
lie on a universal curve 
\begin{equation}
\displaystyle\frac{\Omega_H(r{=}0)}{\Omega}=\text{exp}\Big[-\big(
\frac{\widetilde{H}}{\sigma}\big)^{\alpha}\Big]
\label{Eq:OmegaH}
\end{equation}
(with $\alpha{\simeq}2.8$ and $\sigma$ being the standard deviation), which 
extends over several orders of magnitude. 

\begin{figure}[t]
\centering
\includegraphics[width=0.4\textwidth]{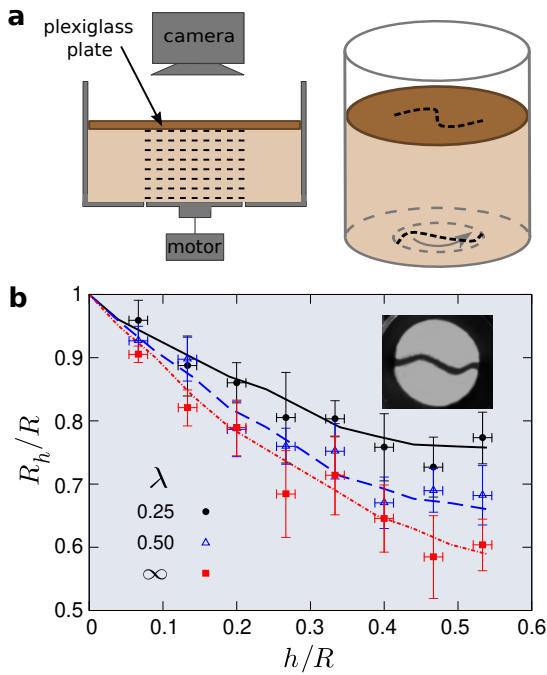}
\caption{Sketch of the experimental setup and evolution of the patterns within 
the bulk. {\bf a}, The split-bottom Couette cell. (left) The dashed lines indicate 
the heights where the initially straight lines of colored beads were created. (right) 
Schematic of the resulting lower and upper patterns after shearing. {\bf b}, Position 
$R_h$ of the shear band at height $h$ within the bulk, for $H{/}R{=}0.55$ and different 
values of $\lambda$. The lines (symbols) represent numerical (experimental) data. To 
achieve the indicated values of $\lambda$ in experiments we apply $P\!\!_\text{ext}
{/}P\!\!_\circ{=}4.0$, $2.0$, or $0$ (corresponding to $\lambda{=}0.25$, $0.50$, 
$\infty$). In simulations we use $g{/}g_\text{earth}{=}0.25,\,0.5,\,1$ and 
$P\!\!_\text{ext}{/}P\!\!_\circ{=}1,\,1,\,0$. The inset shows a typical pattern 
of colored beads within the bulk, obtained experimentally.}
\label{Fig5}
\end{figure}

While the scaling relations\,(\ref{Eq:RHR}),(\ref{Eq:WHR}),(\ref{Eq:OmegaH}) 
for the top-surface characteristics enable one to predict the influence of the 
gravity on the surface flow profile, the variational approach 
accurately determines the shear zone path inside the bulk as well. As the 
final experimental validation of the numerical approach, we first numerically 
obtain the center position of the shear band in the bulk for a given $H$ 
and $P\!\!_\text{ext}$ and several choices of $g$. To achieve the corresponding 
values of $\lambda$ in experiments, we adjust $P\!\!_\text{ext}$ (i.e.\ 
the mass of the top plate). To be able to visualize the bulk flow profile, 
we initially create straight lines of colored beads buried at several heights 
in the bulk and observe the created patterns after shearing by removing the 
upper layers of grains (see Fig.\,\ref{Fig5}a). The numerical and experimental 
results of $R_h$ versus $h$ are compared in Fig.\,\ref{Fig5}b; the agreement 
is excellent, without any adjustable parameter.   

In conclusion, we studied the role of gravity on strain localization in slowly 
sheared granular materials. The variational approach for minimization of energy 
dissipation is computationally a cheaper technique even compared to efficient 
DEM tools for large-scale granular simulations \cite{Shojaaee12,Kloss2012} 
and can be extended to more complex geometries. We obtained universal scaling 
laws describing the characteristics of the surface flow profile, which opens 
the possibility of predicting the shear flow of granular matter in varying 
gravitational environments. Since shear zones mark the regions where energy 
dissipation and catastrophic material failures occur, our results are of 
particular relevance for science and engineering challenges associated with 
planetary and asteroid exploration programs in the coming years and decades. 

\section{Methods}
{\bf Variational approach.} 
In order to perform the numerical simulations based on the variational minimization 
principle, we note that the cylindrical symmetry reduces the problem to an effective 
2D one. A square lattice with the lattice unit being equal to the grain size $d$ is 
considered. Here, the results for $d{/}R{=}10^{-2}$ are presented. An instantaneous 
shear band is obtained via equation~(\ref{Eq:VM1}), which has one lattice unit width. 
Next, the local material strength along the shear band and in its neighborhood are 
randomly updated following the probability distribution $P(\mu^\text{eff})$ reported 
in \cite{Shaebani08}. By repeating the procedure, an ensemble of shear bands are 
obtained from which the characteristics of the shear zone are deduced. 

\noindent{\bf Experimental setup.} 
Our experimental setup is similar in spirit to the modified Couette cell used in 
previous granular shear experiments \cite{Fenistein04,Cheng06,Fenistein06}, but 
with an additional plexiglass plate laid on top of the granular layer to generate 
pressure on the sample; see Fig.\,\ref{Fig5}a. The bottom disk with radius $R{=}45$ 
or $75\,\text{mm}$ is rotated at small angular velocity $\Omega{\simeq}0.1\,\text{rad}
{/}\text{s}$ to avoid rate-dependent stresses \cite{Hartley03}. The mass of the top 
plate ranges from $50$ to $5000\,\text{gr}$ and glass beads with the average diameter 
of $0.5$ to $5\,\text{mm}$ with size polydispersity of about $15\%$ are used to 
prevent ordering of grains near walls \cite{Mueth00,Chambon03}. The gap size at 
the split is less than $400\,\mu\text{m}$ so that no particle can escape. The surface 
flow is monitored from above by a fast CCD camera with pixel resolution $70\,\mu
\text{m}$ and frame rate $30\;\text{s}^{-1}$. The experiments are carried with and 
without the top plate. To obtain the flow profile, particle image velocimetry method 
is used to determine the average angular cross-correlation function for successive 
frames. The axial angular velocity at the top surface is extracted by averaging the 
angular velocities within a small circle of radius $5\,\text{mm}$ around the cylinder 
axis.

JT acknowledges support by Hungarian National Research, Development 
and Innovation Office (NKFIH), under Grant No.\ OTKA K 116036, by the 
BME IE-VIZTKP2020. 

\bibliography{Refs}

\end{document}